# Scientific wealth and inequality within nations

**Gangan Prathap**[1,2]

[1]*Vidya Academy of Science and Technology, Thrissur, Kerala, India 680501.*
*e-mail: gangan@vidyaacademy.ac.in*
[2]*A P J Abdul Kalam Technological University, Thiruvananthapuram, Kerala, India 695016.*
*e-mail: gangan_prathap@hotmail.com*

**Abstract**

We show that the greater the scientific wealth of a nation, the more likely that it will tend to concentrate this excellence in a few premier institutions. That is, great wealth implies great inequality of distribution. The "scientific wealth" is interpreted in terms of citation data harvested by Google Scholar Citations for profiled institutions from all countries in the world.

**Introduction**

The "scientific wealth" of nations is often interpreted in terms of publication and citation data. Early studies along these lines were done by May (1997) and King (2004). Nations with larger R&D investments had larger shares of paper and citation counts (Klavans & Boyack 2017). Indeed, there is a strong relationship between economic and scientific wealth. Leydesdorff & Zhou (2005) further demonstrated that newly emerging powers in science which start from a lower base have relatively higher growth rates. Cole & Phelan (1999) showed that economic forces do not fully account for scientific productivity; social and cultural forces like religion, decentralization and competitiveness were also factors. Cimini *et al.* (2014) use citation data of scientific articles to show that the scientific fitness of each nation, that is, the competitiveness of its research system, depends on the extent to which they diversify as much as possible their research system into as many scientific domains as possible.

So far, no one has looked at how concentration of science output in a few premier institutions within each country (i.e. the inequality in scientific wealth production) is related to the overall scientific wealth of a nation. In this paper, we interpret the "scientific wealth" of a nation in terms of citation data of its various academic institutions harvested by Google Scholar Citations for profiled institutions from all countries in the world. By examining data from three cohorts of countries, we show that the "richer" a country is, the more likely that its scientific excellence will come from a highly concentrated group of premier institutions,

**The Transparent Ranking of Universities**

The Third Edition of TRANSPARENT RANKING: Top Universities by Google Scholar Citations (http://www.webometrics.info/en/node/169) is now available. It uses institutional profiles introduced by Google Scholar Citations (GSC) for providing a ranking of universities using information provided for the groups of scholars sharing the same standardized name and email address of an institution. There are close to one million individual profiles and over 5000 university profiles in GSC. This covers most of the leading academic organizations from nearly 200 countries. The methodology used is described in http://www.webometrics.info/en/node/169. Ranking within each country and globally is done on the basis of descending order of total citations. Since the setting up of a personal profile in GSC is voluntary and some effort is required from each individual to ensure correctness, there will be many errors of omission and commission (i.e. intended or unintended fake, incorrect or duplicate records). Even then, we can have an indicative understanding of the scientific wealth of each country as a count of citations of organizations that make it to the list and also of the unevenness or variance in the distribution of this wealth within a country.

**The methodology of the present exercise and results**

There are 4447 academic institutions in the world which have more than 1000 citations at the time of collection (around 20$^{th}$ December 2016) of TRANSPARENT RANKING. The largest number of institutions is found in the United States of America with 930 institutions (20.9% of the global total). Many small countries have only one institution each and many which do not appear have no institution that makes the cut. The data for China and Russia seem unreliable and in our further exercises these are not considered.

Let us first focus our attention on the records from the United States of America. Let $N$ be the number of institutions that have more than 1000 citations in a country and $C$ be number of total citations. The 930 institutions account for a total of 74,852,741 citations. Note that $N$ is a size-dependent or extensive parameter. The one can think of an average impact term $i = C/N$ as a size-independent measure of the average excellence of the institutions in the country. For the USA, this is 80486.82. Then if $N$ is a zeroth-order measure of performance, $C$ is a first-order measure of scientific output or performance. Following Prathap (2011, 2014), it is possible to define second-order measures of performance such as Exergy $X$ and Energy $E$. The ratio $\eta = X/E$ is a very simple size-independent measure of the degree of unevenness or inequality or of concentration in the distribution. A value of $\eta = 1$ implies absolute equality or evenness of distribution and this is also the default value for this parameter when there is only one institution in the country. For the USA, the corresponding values are $X = 6.02E+12$, $E = 3.02E+13$ and $\eta = 0.200$. That is, excellence is distributed in the USA in a very highly skewed or uneven manner.

In Table 1 we compare the size-dependent and size-indeendent indicators for the world and the United States of America as indexed in TRANSPARENT RANKING. It is seen that the USA maintains an average impact that is nearly twice as high as the global average impact. The global measure of inequality of distribution is higher than that within the USA. That is, globally excellence is concentrated in an even more highly skewed or uneven manner than in the USA.

Following the intuition of Cole & Phelan (1999) that social and cultural forces are significant factors in determining the scientific competitiveness of nations, we look at three cohorts as described in Table 2. Altogether some 52 countries are covered. In one column we have some leading countries as measured by size-dependent measures of performance. China and Russia are omitted from this list as the data from profiled institutions, which in turn depend on the authenticity of data from profiled individuals, seem unreliable. In the second column we look at major Islamic countries (Sarwar & Hassan 2015) to see how social and cultural determinants may affect performance. In the third column we have an agglomeration based on language where the Iberian peninsula countries of Spain and Portugal are taken together with many Latin American countries. In all cases, the nominal GDP measure in billions of US dollars is taken as a measure of the size of the economy. GDP values are those reported by the International Monetary Fund.

Table 3 shows the Pearson's correlation for the size-dependent and size-independent indicators for the 52 countries covered in Table 2. We see a very strong correlation between nominal GDP and the size-dependent research performance indicators. Average impact, $i$, is modestly correlated with GDP; richer countries produce research of higher quality or impact. The size-independent inequality measure is consistently negatively correlated with all the other size-dependent indicators indicators. Figure 1 shows scatter plots illustrating how the size-dependent performance indicators are related to nominal GDP. Indicative lines are also shown with slopes of 1.0, 1.5 and 2.0 respectively. As GDP increases, the scientific perfomance increases, with the higher-order indicators emphasizing the compounding role that impact or quality plays. The zeroth-order indicator, $N$, varies directly with GDP, i.e. richer countries boast of a larger number of institutions that have more than the threshold of 1000 citations. In Figure 2 we have scatter plots showing that the size-independent inequality indicator is negatively correlated with the second-order performance indicators for the three cohorts considered. As nations move towards higher degrees of total excellence, the inequality parameter also increases showing that growth takes place in a concentrated fashion in a few elite institutions.

**Concluding remarks**

We have used citation data harvested by Google Scholar Citations for profiled institutions from all countries in the world as a proxy for the "scientific wealth" of each nation. It is seen that this is very unevenly distributed among the institutions in each country. From correlation analysis and scatter plots we see that the greater the scientific wealth of a nation the more likely is it that it will tend to concentrate this excellence in a few premier institutions. That is, great wealth implies great inequality of distribution.

Table 1. The size-dependent and size-independent indicators for the world and the United States of America.

| Indicator | WORLD | USA | USA as percentage of World |
|---|---:|---:|---:|
| $N$ | 4447 | 930 | 20.9 |
| $C$ | 190531311 | 74852741 | 39.3 |
| $i$ | 42844.91 | 80486.82 | - |
| $X$ | 8.16E+12 | 6.02E+12 | 73.8 |
| $E$ | 5.14E+13 | 3.02E+13 | 58.7 |
| $\eta$ | 0.159 | 0.200 | - |

Table 2. Three cohorts taken up for examining the nature of relationships between size-dependent and size-independent indicators for various countries.

| Top 12 | Islamic | Iberia & Latin America |
|---|---|---|
| USA | Algeria | Argentina |
| UK | Bahrain | Bolivia |
| Canada | Bangladesh | Brazil |
| Italy | Brunei Darussalam | Chile |
| South Korea | Egypt | Colombia |
| Germany | Indonesia | Costa Rica |
| Spain | Iran | Cuba |
| France | Iraq | Ecuador |
| Japan | Jordan | Gautemala |
| Brazil | Kazakhstan | Honduras |
| India | Kuwait | Mexico |
| Portugal | Lebanon | Panama |
| | Libya | Paraguay |
| | Malaysia | Peru |
| | Morocco | Portugal |
| | Oman | Spain |
| | Pakistan | Uruguay |
| | Palestine | Venezuela |
| | Qatar | |
| | Saudi Arabia | |
| | Sudan | |
| | Syria | |
| | Tunisia | |
| | Turkey | |
| | United Arab Emirates | |

Table 3. Pearson's correlation for the size-dependent and size-independent indicators for the 52 countries covered in Table 2.

| Pearson's correlation | N | C | X | E | GDP $b | i | η |
|---|---|---|---|---|---|---|---|
| N | 1.00 | 0.91 | 0.87 | 0.88 | 0.95 | 0.43 | -0.38 |
| C | 0.91 | 1.00 | 0.99 | 0.99 | 0.97 | 0.55 | -0.25 |
| X | 0.87 | 0.99 | 1.00 | 0.99 | 0.94 | 0.59 | -0.23 |
| E | 0.88 | 0.99 | 0.99 | 1.00 | 0.95 | 0.48 | -0.21 |
| GDP $b | 0.95 | 0.97 | 0.94 | 0.95 | 1.00 | 0.49 | -0.31 |
| i | 0.43 | 0.55 | 0.59 | 0.48 | 0.49 | 1.00 | -0.34 |
| η | -0.38 | -0.25 | -0.23 | -0.21 | -0.31 | -0.34 | 1.00 |

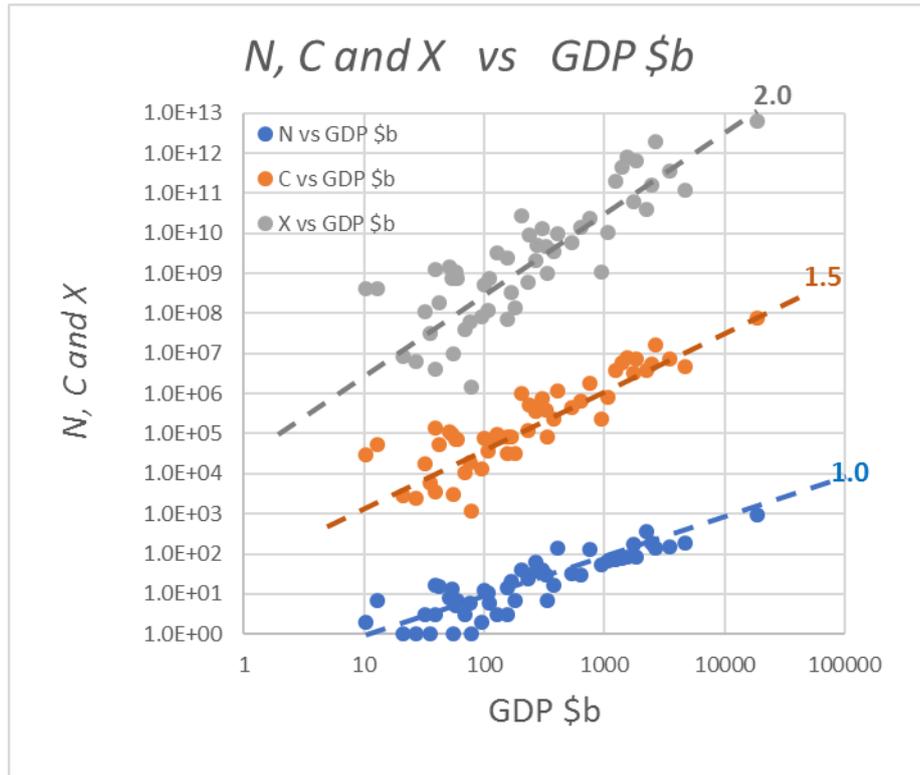

Figure 1. Scatter plots showing how the size-dependent performance indicators are related to nominal GDP

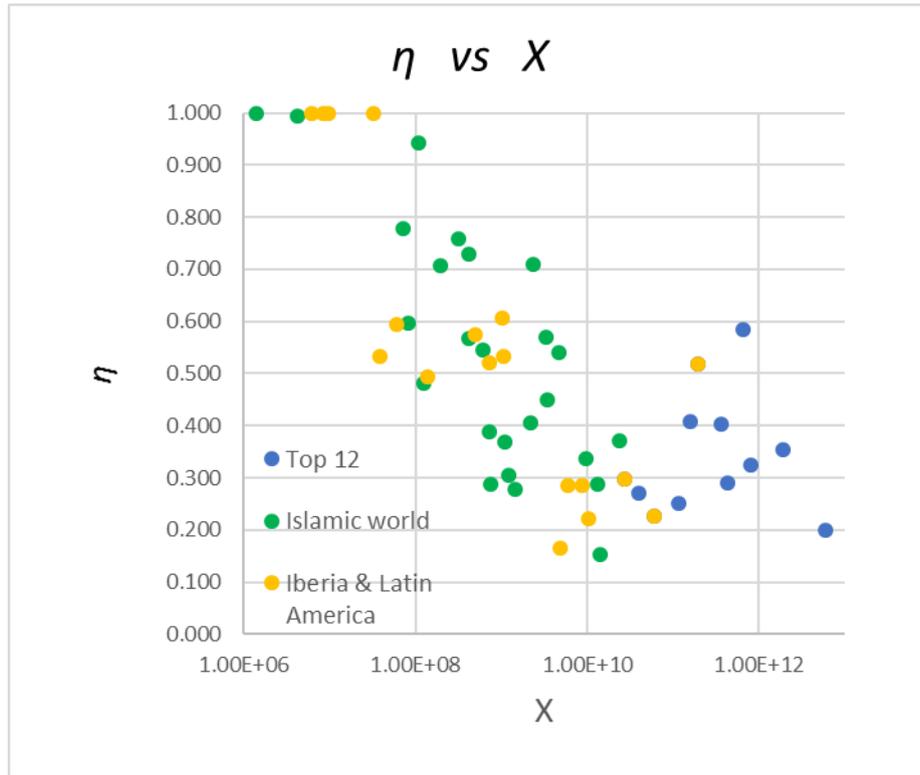

Figure 2. Scatter plots showing how the size-independent inequality indicator is negatively correlated with the second-order performance indicators for the three cohorts considered.